# Fundamental Issues and Problems in the Realization of Memristors

Paul Meuffels[*,1] and Rohit Soni[2]

[1]*Forschungzentrum Jülich GmbH, Peter Grünberg Institut, D-52425 Jülich, Germany*

[2]*Nanoelektronik, Technische Fakultät, Christian-Albrechts-Universität zu Kiel, D-24143 Kiel, Germany*

**Abstract**

In 2008, researchers at the Hewlett-Packard (HP) laboratories claimed to have found an analytical physical model for a genuine memristor device [1]. The model is considered for a thin $TiO_2$ film containing a region which is highly self-doped with oxygen vacancies and a region which is less doped, *i.e.*, a single-phase material with a built-in chemical inhomogeneity sandwiched between two platinum electrodes. On base of the proposed model, Strukov *et al.* [1] were able to obtain the characteristic dynamical state equation and current-voltage relation for a genuine memristor. However, some fundamental facts of electrochemistry have been overlooked by the authors while putting forward their model, namely the coupling of diffusion currents at the boundary between both regions. The device will operate for a certain time like a "chemical capacitor" until the chemical inhomogeneity is balanced out, thus violating the essential requirement on a genuine memristor, the so-called "no energy discharge property". Moreover, the dynamical state equation for the HP-memristor device must fail as this relation violates by itself Landauer's principle of the minimum energy costs for information processing. Maybe, such an approach might be upheld if one introduces an additional prerequisite by specifying the minimum amount of electric power input to the device which is required to continuously change internal, physical states of the considered system. However, we have reasonable doubts with regard to this.

**Keywords**: memristor, electrochemistry, polarization voltage, chemical capacitor, thermodynamics, Landauer's principle

[*]Corresponding author: p.meuffels@fz-juelich.de

# 1. Introduction

In 2008, Strukov *et al.* [1] published a paper in the journal *Nature* entitled "*The missing memristor found*". Such a finding would be a unique breakthrough because one could then realize a working "fourth fundamental circuit element". The "memristor concept" was theoretically introduced by Leon Chua [2] on base of symmetry arguments looking at all possible relations between the four circuit variables current $I$, voltage $V$, charge $Q$ and magnetic flux $\phi$. A memristor device would exhibit peculiar current-voltage characteristics which would be completely different from those exhibited by resistors, capacitors and inductors [2, 3]: The device's resistance, therefore called memristance, would "remember" the charge that has flowed through the device. Memristors would be passive circuit elements and their behavior could in no way simulated by means of resistor-capacitor-inductor networks. In this sense, the memristor would be a real "fourth fundamental circuit element" the properties of which could lead to a number of new applications [3, 4].

Since the announcement [1] that the missing memristor had been discovered by researchers at the Hewlett–Packard laboratories, both the "memristor concept" and the HP-memristor device came into the focus of ongoing intense discussions (see, for example, [3, 5 - 9]). Nowadays, it seems that the theoretical "memristor concept" might indeed be realized as a functioning device in physical reality, despite there are some serious doubts [9]. This belief is mainly based on the "physical model" for a memristor device which was presented by Strukov *et al.* [1] in their letter to *Nature*. It is claimed that memristance effects arise naturally in nanoscopic systems owing to the coupling of ionic and electronic transport under external bias voltage.

We have some severe questions regarding the HP-memristor model, especially, when viewing it under the aspects of electrochemistry. It is thus necessary to analyze the HP-memristor model in more detail as will be done in chapter 3 and 4. We will show that the model suffers from a severe inconsistency as it ignores diffusion related effects on the current transport properties of the device. In chapter 5, we will furthermore demonstrate on base of thermodynamic considerations that the HP-memristor model must fail because it violates Landauer's principle of the minimum energy costs for information processing. However, before starting the analysis, the essential features of the "memristor concept" will be discussed at first.

# 2. The memristor concept

Three two-terminal circuit elements are well known via three constitutive relations $F$ between the circuit variables $I$ and $V$, $Q$ and $V$ and $I$ and $\phi$ : $\boldsymbol{F}_{\text{resistor}}(I,V) = 0$ for the resistor, $\boldsymbol{F}_{\text{capacitor}}(Q,V) = 0$ for the capacitor and $\boldsymbol{F}_{\text{inductor}}(\phi,I) = 0$ for the inductor. As theoretically proposed by Leon Chua [2], there might be a fourth two-terminal circuit element, namely the memristor, defined by a constitutive relation between the charge $Q$ that has flowed in a system and the magnetic flux $\phi$: $\boldsymbol{F}_{\text{memristor}}(\phi,Q) = 0$. Using this constitutive relation, a memristor is said to be charge-controlled (flux-controlled) if this relation can be expressed as a single valued function of the charge $Q$ (flux $\phi$) [2]. For the voltage applied across, for example, a charge-controlled memristor one gets

$$V(t) = R\big(Q(t)\big) * I(t) \tag{1a}$$

$$\frac{dQ(t)}{dt} = I(t) \, , \tag{1b}$$

where $t$ is the time, $V(t)$ the time dependent voltage applied at the two terminals of the memristor and $I(t)$ the time dependent current flowing through the memristor. $R(Q(t))$ is called the incremental memristance as the physical unit of $R$ is [V/A]. The value of $R$ at a given time $t$ depends thus only on the electrical charge $Q$ that has flowed through the memristor up to time $t$. To say it in other words, the memristor memorizes the charge that has run through it.

Strukov *et al*. [1] have pointed out that no one has been able to come up with a realistic physical model that satisfies the set of equations (1a) and (1b). Thus, the memristor might be merely a concept which cannot be realized in our world. However, one can generalize the concept by giving up the idea that there is a direct functional relation between the memristance $R$ and the charge $Q$. By this means, one gets rid of the connotation between the total electrical charge $Q(t)$ that has been transported through a system up to time $t$ and the magnetic flux $\phi(t)$ established somewhere in the system at time $t$.

Equation (1a) can be expressed in a more general form if one assumes that the memristance $R$ depends on a dynamical state variable $w$ of the system which itself evolves in time according to some prescribed differential equation. In case the time dependence of $w$ is related to the current $I(t)$ that is flowing through the system, this generalization allows to define a broader class of nonlinear dynamical systems which have been labeled memristive systems [3, 10]. Referring to [3], the resistance *vs*. state map of a memristive system obeys Ohm's Law, except that the resistance $R$ is not a constant. A memristive system can thus generally be described by the following set of mathematical equations [3]:

State dependent Ohm's law of a memristive system:

$$V(t) = R\big(w(t), I(t)\big) I(t) \tag{2a}$$

Dynamical state equation of a memristive system:

$$\frac{d\,w(t)}{dt} = f(w(t), I(t)) \tag{2b}$$

$w(t)$ is the internal dynamical state variable of the device and $R(w(t), I(t))$ is a generalized memristance depending on the state of the device. If the memristance $R(w(t), I(t))$ is positive at any given time, then the total electric energy $E_{\text{electric}}$ fed in to a memristor between time $t_1$ and $t_2$

$$E_{\text{electric}} = \int_{t_1}^{t_2} V(t')I(t')dt' = \int_{t_1}^{t_2} R(w(t'), I(t')) * (I(t'))^2 \, dt' \geq 0 \tag{3}$$

is always positive or zero [5]. Thus, memristor devices are passive devices and seem to act like real resistors: The passage of an electric current through the memristor releases Joule heat which is then dissipated by heat exchange to the environment [5]. There is another important feature as noted by Pershin and Di Ventra [5]: When $V(t) \equiv 0$, then $I(t) \equiv 0$ (and vice versa). Pershin and Di Ventra [5] called this feature the "no energy discharge property", *i.e.*, a memristor device does not store energy like a capacitor or an inductor.

We have now advanced a little bit as the new "memristor concept" (equation (2a) and (2b)) seems to be more realistic than the older one (equation (1a) and (1b)). Nevertheless, there is still an open question: "Can such a memristor be realized as a functioning device in physical reality?". One has thus to propose a physical model for a real memristor device which satisfies the dynamical state equations (2a) and (2b) but does not violate fundamental laws of physics.

### 3. The HP-memristor model

In the above mentioned paper published by the HP-group [1], the authors argue: "*Here we show, using a simple analytical example, that memristance arises naturally in nanoscale systems in which solid-state electronic and ionic transport are coupled under an external bias. These results serve as the foundation for understanding a wide range of hysteretic current-voltage behavior observed in many nanoscale devices.*". As this is indeed a far-reaching statement, there is need for a closer look at this analytical example under aspects of electrochemistry.

The physical model for the HP-memristor device is schematically shown in Fig 1. The model is considered for a titanium dioxide ($TiO_2$) film of thickness $L$ which is sandwiched between two platinum contacts. The film itself consists of two regions with different oxygen deficiencies: The first region which is labeled "ON region" consists of $TiO_{2-x_{ON}}$ with a high oxygen deficiency $x_{ON}$, whereas the second region which is labeled "OFF region" consists of nearly stoichiometric $TiO_{2-x_{OFF}}$ with a very low oxygen deficiency $x_{OFF} \ll x_{ON}$. We have therefore a single-phase material with a built-in chemical inhomogeneity which is sandwiched between two metal electrodes.

We assume now that deviations from stoichiometry in $TiO_2$ are accommodated by oxygen vacancies on the oxygen sublattice. The neutral oxygen vacancies $V_O^x$ can act as donor-type dopants (self-doping) according to the ionization reaction (written in Kröger and Vink notation [11]):

$$V_O^x \Leftrightarrow V_O^{\bullet\bullet} + 2e' \tag{4a}$$

$$K_{ion} = \frac{c_{V_O^x}}{c_{V_O^{\bullet\bullet}} (c_{e'})^2} \tag{4b}$$

$V_O^{\bullet\bullet}$ denotes a double ionized oxygen vacancy with a net charge of $+2q$ and $e'$ a "free" electron with a net charge of $-q$ ($q$ is the elementary charge). $K_{ion}$ is the equilibrium constant for reaction (4a) following from the law of mass action and $c_i$ is the concentration of defect species i with unit [cm$^{-3}$]. We assume for the sake of simplicity that the charged oxygen vacancies are the majority ionic defects and the electrons the majority electronic defects in the considered system so that the charge neutrality condition reads approximately $2c_{e'} \cong c_{V_O^{\bullet\bullet}}$.

Owing to the large difference in oxygen stoichiometry ($x_{OFF} \ll x_{ON}$), the self-doped ON region will have a high electronic conductivity whereas the nearly stoichiometric OFF region will have a much lower electronic conductivity compared to that in the ON region. Presuming that the oxygen vacancies in the ON and OFF region are fully ionized ($K_{ion} \rightarrow 0$) and uniformly distributed over the corresponding region, the concentration of the charged oxygen vacancies can be related to the oxygen deficiency $x$ in $TiO_{2-x}$ via

$$c_{V_O^{\bullet\bullet},ON} \cong x_{ON}/V_{uc} \tag{5a}$$

$$c_{V_O^{\bullet\bullet},OFF} \cong x_{OFF}/V_{uc} , \tag{5b}$$

where $V_{uc}$ is the unit cell volume of $TiO_{2-x}$. Let us now mark the electrons ($e'$, valence -1) with index n and the charged ionic defect species ($V_O^{\bullet\bullet}$, valence +2) with index V. The electrons and ionic defects should have average mobilities, $\beta_n$ and $\beta_V$, respectively. Generally, as electrons are considerably more mobile than ionic defects, one has $\beta_n \gg \beta_V$. Denoting the initial spatial coordinate $x$ of the boundary between both regions with $w_0$, we have for the self-doped ON region ($0 \leq x \leq w_0$):

$$c_{V,ON} \cong x_{ON}/V_{uc} \tag{6a}$$

$$c_{n,ON} \cong 2c_{V,ON} \tag{6b}$$

$$\sigma_{ON} = q\beta_n c_{n,ON} + 2q\beta_V c_{V,ON} \cong q\beta_n c_{n,ON} \tag{6c}$$

and for the nearly stoichiometric OFF region ($w_0 \leq x \leq L$):

$$c_{V,OFF} \cong x_{OFF}/V_{uc} \quad \ll c_{V,ON} \tag{7a}$$

$$c_{n,OFF} \cong 2c_{V,OFF} \quad \ll c_{n,ON} \tag{7b}$$

$$\sigma_{OFF} = q\beta_n c_{n,OFF} + 2q\beta_V c_{V,OFF} \cong q\beta_n c_{n,OFF} \quad \ll \sigma_{ON} \tag{7c}$$

$L$ is the thickness of the film with unit [cm], $\beta_n$ and $\beta_V$ ($\ll \beta_n$) are the mobilities of the electronic and ionic defects, respectively, with unit [cm$^2$ V$^{-1}$ sec$^{-1}$] and $\sigma_{ON}$ and $\sigma_{OFF}$ are the electric conductivities in the ON and OFF region, respectively, with unit [A cm$^{-1}$ V$^{-1}$].

According to Strukov *et al*. [1], applying an external bias $v(t)$ across the device will now move the boundary between the ON and OFF region by causing the charged ionic defects to drift, *i.e.*, the boundary can be displaced in a bidirectional way by means of an external voltage stress. To arrive at the equations which are presented in [1], there is requirement for an additional side condition: The relevant concentration of the charged oxygen vacancies in the ON region (see equation (6a)) must not change essentially in course of time despite the moving boundary. Starting with the simple case of ohmic electronic conduction in both regions and linear ionic drift in a uniform electric field in the ON region (and keeping the side condition in mind), one can now derive the current-voltage, $i(t)$-$v(t)$, relation for the HP-memristor device. The electric fields in both regions are related to the total electric current density $i(t)$ (with unit [A cm$^{-2}$]) flowing through the device by

$$E_{ON}(t) = \frac{i(t)}{\sigma_{ON}} \quad \text{for the ON region} \quad 0 \leq x \leq w(t) \tag{8a}$$

$$E_{OFF}(t) = \frac{i(t)}{\sigma_{OFF}} \quad \text{for the OFF region} \quad w(t) \leq x \leq L , \tag{8b}$$

where $w(t)$ denotes now the time dependent spatial coordinate of the boundary between both regions. The voltage $v(t)$ (at $x = L$, defined with respect to the reference point "ground" at $x = 0$, see Figure 1) is now given by

$$v(t) = -\Big(E_{ON}(t)w(t) + E_{OFF}(t)\big(L - w(t)\big)\Big), \tag{9}$$

and one arrives at the state dependent Ohm's law of the HP-memristor device

$$v(t) = -\left(\frac{1}{\sigma_{ON}}w(t) + \frac{1}{\sigma_{OFF}}\big(L - w(t)\big)\right)i(t) \tag{10a}$$

or

$$v(t) = -\left(R_{ON}\frac{w(t)}{L} + R_{OFF}\left(1 - \frac{w(t)}{L}\right)\right)i(t), \tag{10b}$$

with defining $R_{ON} = L/\sigma_{ON}$ and $R_{OFF} = L/\sigma_{OFF}$.

According to Strukov *et al*. [1], the positively charged ionic defects should now linearly drift under the uniform electric field in the ON region. Here is the first problematic point which is not discussed in [1]: What is the electric field strength at the boundary $w(t)$ in case on has two limiting values $E_{ON}(t)$ and $E_{OFF}(t)$? Such a discontinuity would indicate a very unstable physical condition at the boundary. However, continuing with the argumentation in [1], one should have for the ionic current density

$$i_{V,ON} = +2q\beta_V c_{V,ON} E_{ON}(t) = +2q\beta_V c_{V,ON} \frac{i(t)}{\sigma_{ON}} , \qquad (11)$$

where $i_{V,ON}$ is the electric current density of the charged oxygen vacancies with unit [A cm$^{-2}$]. For the drift velocity, one would obtain

$$v_{V,ON}^{drift} = \frac{i_{V,ON}}{+2qc_{V,ON}} = \beta_V \frac{i(t)}{\sigma_{ON}} . \qquad (12)$$

$w(t)$ would thus change within an infinitesimal time interval $dt$ according to the dynamical state equation of the HP-memristor device

$$\frac{dw(t)}{dt} = v_{V,ON}^{drift} = \beta_V \frac{1}{\sigma_{ON}} i(t), \qquad (13a)$$

or

$$\frac{dw(t)}{dt} = \beta_V \frac{R_{ON}}{L} i(t) = \gamma i(t) . \qquad (13b)$$

The model for the HP-memristor device leads thus to the state equations of the simplest memristive system one can think of: The internal dynamical state variable $w(t)$ is simply proportional to $\int_{-\infty}^{t} i(t')dt'$.

## 4. Analysis of the HP-memristor model under aspects of electrochemistry

At first glance, the simple analytical model for the HP-memristor device seems indeed to work, at least approximately. At second glance, however, one detects a severe inconsistency. This inconsistency merely arises because Strukov *et al.* [1] did not consider a slight, but important detail, namely the coupling of the diffusive currents of the electrons and oxygen vacancies at the boundary between both regions. A single-phase system with a "built-in" chemical inhomogeneity as sketched in Fig. 1 tends to re-equilibrate under all circumstances, *i.e.*, as long as an oxygen vacancy concentration gradient exists inside the material one is concerned with the diffusion of the involved defect species.

Let us assume for the sake of simplicity that the concentrations of all defect species are so low that one can treat the system within the framework of dilute solid solutions. Using the Nernst-Planck equations, the particle flux densities for the electronic and ionic defects can then be written as (see, for example, [12 - 15]):

$$j_n(x,t) = -D_n \frac{\partial c_n(x,t)}{\partial x} - \beta_n c_n(x,t) E(x,t) \qquad (14a)$$

$$j_V(x,t) = -D_V \frac{\partial c_V(x,t)}{\partial x} + \beta_V c_V(x,t) E(x,t) . \qquad (14b)$$

$j_n(x,t)$ and $j_V(x,t)$ denote the particle flux densities of the electrons and charged oxygen vacancies, respectively, with unit [cm$^{-2}$ sec$^{-1}$] and $D_n$ and $D_V$ are the self diffusion coefficients of the electrons and charged oxygen vacancies, respectively, with unit [cm$^2$ sec$^{-1}$]). The first terms in equations (14a) or (14b) are the diffusion fluxes which are proportional to the concentration gradients at a given position *x*, while the second terms are the drift fluxes which are proportional to the concentration and the electrical field $E(x,t)$ at *x*. The mobility of each defect species is related to the self diffusion coefficient by means of the Nernst-Einstein relation,

$$\beta_n = \frac{qD_n}{k_B T} \tag{15a}$$

$$\beta_V = \frac{2qD_V}{k_B T}, \tag{15b}$$

where $T$ is the temperature and $k_B$ the Boltzmann constant.

According to equations (14a) and (14b), particle fluxes are always established around a chemical gradient even when there is no externally applied field. Let us now assume that no external electric field is applied to our device. In order to degrade the steep concentration gradient, the majority charge carriers in the ON region, *i.e.*, electrons and charged oxygen vacancies, will start to diffuse from the ON to the OFF region. As the extremely mobile electrons do not care how fast the ionic defects are, the system goes immediately for establishing a uniform electrochemical potential $\eta_n$ (Fermi energy) of the electrons. Because the more mobile electrons are always "in advance" of the less mobile ionic defects, an electric potential difference develops between the ON and OFF region arising from a slight misadjustement of the charge balance at the diffusion front. An attractive electrostatic field is established between both defect ensembles which acts to accelerate the less mobile defects and to decelerate the more mobile ones.

If the concentration misadjustment is very small, the assumption of local "quasi electroneutrality" at the diffusion front can be made, *i.e.*, in all considerations other than those concerned with space charge regions the concentrations (in terms of equivalents) may be taken as equal [16],

$$c_n(x,t) = 2c_V(x,t). \tag{16}$$

To maintain local "quasi electroneutrality", the particle flux density of one defect species must be finally matched by an equivalent particle flux density of the other species, *i.e.*,

$$j_n(x,t) = 2\, j_V(x,t). \tag{17}$$

This means that the defects are forced to flow together from the ON to the OFF region so that they have their diffusion coefficients "in common". This gives rise to the so-called ambipolar or chemical diffusion [16 - 18] which – in a figurative sense – is the diffusion of a "electroneutral defect pair" or building unit $|V_O^{\bullet\bullet} - 2e'|$. The individual fluxes can now be written in terms of a chemical diffusion coefficient $\widetilde{D}$ according to

$$j_n(x,t) = -\widetilde{D}\, \frac{\partial c_n(x,t)}{\partial x} \tag{18a}$$

$$j_V(x,t) = -\widetilde{D}\, \frac{\partial c_V(x,t)}{\partial x}. \tag{18b}$$

Thus, under the aspect of local "quasi electroneutrality", the concentration gradients of the coupled defect species must obey:

$$2\frac{\partial c_V(x,t)}{\partial x} = \frac{\partial c_n(x,t)}{\partial x}. \tag{19}$$

To describe the electric transport properties of a mixed ionic electronic conducting (MIEC) material when electronic and ionic currents are coupled both under an external voltage bias and a chemical concentration gradient, we follow now the elegant paper by Isaaki Yokota [19] and start with writing down the electrochemical potentials of the electronic and ionic defects:

$$\eta_n(x,t) = \mu_n^0 + k_B T\, ln(c_n(x,t)) - q\varphi(x,t) \tag{20a}$$

$$\eta_V(x,t) = \mu_V^0 + k_B T\, ln(c_V(x,t)) + 2q\varphi(x,t). \tag{20b}$$

$\eta_n(x,t)$ and $\eta_V(x,t)$ are the electrochemical potentials and $\mu_n^0$ and $\mu_V^0$ are the standard chemical potentials of the electronic and ionic defects, respectively, and $\varphi$ is the electric potential. The electronic and ionic current densities can now be expressed, respectively, as

$$i_n(x,t) = \sigma_n(x,t)\frac{\partial(\eta_n(x,t)/q)}{\partial x}$$
$$= \sigma_n(x,t)\frac{\partial(\mu(x,t)/(2q))}{\partial x} - \sigma_n(x,t)\frac{\partial(\eta_V(x,t)/(2q))}{\partial x} \tag{21a}$$

$$i_V(x,t) = -\sigma_V(x,t)\frac{\partial(\eta_V(x,t)/(2q))}{\partial x}, \tag{21b}$$

where $\sigma_n(x,t)$ and $\sigma_V(x,t)$ are the electronic and ionic conductivities, respectively, and $\mu(x,t)$ is the formally defined chemical potential of the neutral building unit $|V_O^{\bullet\bullet} - 2e'|$ according to

$$\mu(x,t) = \eta_V(x,t) + 2\eta_n(x,t). \tag{22}$$

The total electric current density $i(t)$ flowing through the system at time $t$ is equal to the sum of the local electronic and ionic current densities at every position $x$:

$$i(t) = i_n(x,t) + i_V(x,t)$$
$$= \sigma_n(x,t)\frac{\partial(\mu(x,t)/(2q))}{\partial x} - \sigma(x,t)\frac{\partial(\eta_V(x,t)/(2q))}{\partial x}, \tag{23}$$

where $\sigma(x,t) = \sigma_n(x,t) + \sigma_V(x,t)$ is the total electric conductivity. Upon substituting this result, one can eliminate $\eta_V$ from equations (21a) and (21b) and obtains:

$$i_n(x,t) = \frac{\sigma_n(x,t)}{\sigma(x,t)}i(t) + \frac{\sigma_n(x,t)\sigma_V(x,t)}{\sigma(x,t)}\frac{\partial(\mu(x,t)/(2q))}{\partial x} \tag{24a}$$

$$i_V(x,t) = \frac{\sigma_V(x,t)}{\sigma(x,t)}i(t) - \frac{\sigma_n(x,t)\sigma_V(x,t)}{\sigma(x,t)}\frac{\partial(\mu(x,t)/(2q))}{\partial x}. \tag{24b}$$

Here, we have now arrived at the essential point for a correct analytical description of a MIEC system as sketched in Figure 1. As we use metal electrodes to contact our MIEC, the electric potential difference or voltage drop between both electrodes is given by

$$v(t) = \varphi(L,t) - \varphi(0,t) = \frac{\eta_n(L,t) - \eta_n(0,t)}{-q}, \tag{25}$$

which is, according to equations (21a) and (24a), equal to

$$v(t) = -\int_0^L \frac{i(t)}{\sigma(x',t)}dx' - \int_0^L \frac{\sigma_V(x',t)}{\sigma(x',t)}\frac{\partial(\mu(x',t)/(2q))}{\partial x'}dx'. \tag{26}$$

This relation has been overlooked by Strukov *et al.* [1] while forming their analytical physical model. The measured voltage between the metal contacts obeys equation (26) where the first term is of ohmic nature, while the second term is the so-called "polarization potential" or "diffusion potential" due to concentration polarization effects [18, 19]. Thus, equations (24a), (24b) and (26) are the only valid relations which can be derived on base of analytical models for systems like the HP-memristor device.

To relate the ionic defect current $i_V(x,t)$ and the voltage $v(t)$ to the concentration gradient of the charged oxygen vacancies, one can now make use of the assumption that local "quasi-electroneutrality" prevails in the system and arrives with the help of equations (16) and (19) at

$$v(t) \cong -i(t)\int_0^L \frac{1}{\sigma(x',t)}dx' - 2q\widetilde{D}\int_0^L \frac{1}{\sigma_n(x',t)}\frac{\partial c_V(x',t)}{\partial x'}dx', \tag{27}$$

and

$$i_V(x,t) \cong \frac{\sigma_V(x,t)}{\sigma(x,t)} i(t) - 2q\widetilde{D} \frac{\partial c_V(x,t)}{\partial x} , \qquad (28)$$

where $\widetilde{D} = 3D_n D_V/(D_n + 2D_V)$ is the chemical diffusion coefficient. These equations for the voltage $v(t)$ (equation (27)) and ionic defect current $i_V(x,t)$ (equation (28)) differ essentially from those which were obtained by Strukov *et al.* [1] on base of their analytical model (see equation (10a) and (11), respectively). A comparison reveals that the HP-memristor model renders only the "ohmic" terms on the right sides of equation (27) and (28). Even regarded as an approximation, the HP-memristor model fails under all circumstances where the ionic defects, here the charged oxygen vacancies, have a – maybe very low –, but finite mobility $\beta_V > 0$ or diffusion coefficient $D_V = k_B T \beta_V / 2q > 0$, *i.e.*, the HP-memristor model would be valid only in case $D_V \equiv 0$.

The boundary between the ON and OFF region will always "move", just as long as a chemical diffusion flux is established in the system owing to the oxygen vacancy concentration gradient. By using equation (28) and

$$\sigma_V(x,t) = 4q^2 c_V(x,t) \frac{D_V}{k_B T} , \qquad (29)$$

one obtains for the drift velocity of the charged oxygen defects

$$v_V^{\text{drift}}(x,t) = \frac{1}{2qc_V(x,t)} \left\{ \frac{4q^2 c_V(x,t) D_V}{k_B T \sigma(x,t)} i(t) - 2q\widetilde{D} \frac{\partial c_V(x,t)}{\partial x} \right\} . \qquad (30)$$

The "diffusion velocity" of the boundary depends on the magnitude of the chemical diffusion coefficient and on the steepness of the concentration gradient. Thus, one will never arrive at a dynamical state equation like equation (13a). One is always faced with a generalized thermodynamic force which drives the system back to its absolute stable equilibrium state where all defect species are uniformly distributed over the MIEC. With respect to information processing, such a device would simply erase all stored information by itself as it always strives for reaching equilibrium.

Moreover, as pointed out by Pershin and Di Ventra [5], a memristor device must exhibit the "no energy discharge property", *i.e.*, one has always $v(t) \equiv 0$ when $i(t) \equiv 0$ or vice versa. However, as can be seen from equation (27), the device will exhibit a maybe low, but nevertheless finite depolarization current density $i(t)$ under "closed circuit" conditions ($v(t) \equiv 0$) or a finite "polarization voltage" or "diffusion voltage" $v(t)$ under "open circuit" conditions ($i(t) \equiv 0$), as long as a chemical concentration gradient exists somewhere inside the device material. Viewed under this aspect, the HP-memristor device would operate like a – so to speak – "chemical capacitor" until the system has arrived at its equilibrium state. Afterwards, depending on the boundary conditions, the system will merely act in the manner of a Maxwell-Wagner [15, 18, 19] or Hebb-Wagner polarization cell [18, 19]. Thus, following the correct analytical description, the HP-memristor device would violate the "no energy discharge property" of a genuine memristor device as it is able to perform a certain amount of work in course of time.

One could now argue that the present analysis is related to systems of macroscopic dimensions. That is the case, of course, because the original HP-memristor model is designed for macroscopic systems, too. The assumptions of ohmic electronic conduction and linear ionic drift in an uniform electric field indirectly imply that we are concerned with a macroscopic system where the electric transport properties are mainly determined by the bulk properties of the MIEC. At the nanoscale, the HP-memristor model would completely collapse. With decreasing dimensions of a device, the conditions existing at the interface between the metal electrodes and the MIEC (Fermi level alignment) take control of the overall electric conduction process [20]. One ends soon up in so-called "flatband" situations, where the electric transport properties are mainly ruled by the difference between the work function of the electrode's metal and the electron affinity of the semiconducting MIEC. In order to calculate the current-voltage characteristics of such devices, one has to rely on numerical methods [12, 15].

## 5. Thermodynamic constraints for the realization of memristor devices

In the preceding chapter we have shown that the HP-memristor model is physically invalid when one is considered with the behavior of a real MIEC device. We take now the chance to think about the HP-memristor model in the framework of thermodynamics in order to explain why such models must fail. Let us assume in some kind of "gedankenexperiment" that we have found a device which follows the memristor state equations

$$V(t) = R(w(t))I(t) \tag{31a}$$

$$\frac{dw(t)}{dt} = \gamma I(t). \tag{31b}$$

We further assume that the memristance of the device is $R(w(t))$ at time $t$, *i.e.*, the device should be in the internal memristor state $\{w(t)\}$ at time $t$ with respect to the internal state variable $w(t)$. After an infinitesimal time interval of length $dt$, the internal state variable $w(t)$ will thus be altered according to

$$w(t + \delta t) = w(t) + \gamma I(t)dt. \tag{32}$$

Correspondingly, the device memristance has changed from $R(w(t))$ to a value of $R(w(t + dt))$, *i.e.*, the device is now in the internal state $\{w(t + dt)\}$. Both states must be real states in a thermodynamic sense as they must be correlated to some real physical modification of the system, *i.e.*, in case of the HP-memristor, for example, the change in the extensiveness of the self-doped region in $TiO_2$. Otherwise, one would observe no real change in the device resistance. Furthermore, according to the dynamical state equation (31b), these states will not alter with time when no more current is applied to the system, *i.e.*, after switching off the current $I(t)$ at time $t$ or $t + dt$ the system would remain in state $\{w(t)\}$ or $\{w(t + dt)\}$, respectively. Thus, both memristor states must be some type of equilibrium states for the system, again in a thermodynamic sense. This is the physical meaning when saying that a memristor device remembers the amount of charge that has flowed through it; with respect to the thermodynamics of information processing, the memristor stores some "information" when it resides in its memristor states $\{w(t)\}$ or $\{w(t + dt)\}$.

As our hypothetical memristor should have a correspondent in physical reality, it can be analyzed in terms of thermodynamics. Let us assume that we are operating under conditions of fixed temperature $T$ and pressure $p$. The relevant thermodynamic state function for the device would be the Gibbs free energy $G$. If one accepts now that the internal memristor states represent some kind of equilibrium states of the device, these states must be separated from each other by a finite free energy barrier. Equilibrium states are attractor states, *i.e.*, with respect to the Gibbs free energy we must always have $dG(w(t)) = 0$ and $dG(w(t + dt)) = 0$ and, additionally, $d^2G(w(t)) > 0$ and $d^2G(w(t + dt)) > 0$ [21]. Otherwise, one would be faced with an indifferent situation and the device would arbitrarily fluctuate from one state to another so that one ends up with nothing else but a very noisy resistor.

That means, however, that the internal memristor or "information" states must be separated from each other by Gibbs free energy barriers. To transfer the system from one equilibrium state to another, one has to perform real work on the system in order to surmount these Gibbs free energy barriers. Any "directed" perturbation of an equilibrium state requires the input of some generalized "thermodynamic work" to counteract the restoring "thermodynamic force" which tends to drive back the system to the equilibrium state. According to Landauer's principle [22 - 24], on average, at least an amount of $RTln2$ ($R$ is the molar gas constant) of work is required to erase some "information" from a memory device. Regarding our hypothetical memristor device, we have thus to perform at least this amount of work on our device to accomplish the transition from state $\{w(t)\}$ to state $\{w(t + dt)\}$ because we have to "erase" the information stored in state $\{w(t)\}$.

As we feed in electric energy to our device, the maximum amount of work $W_{\max}$ which can be delivered to the system within time interval $dt$ will be

$$W_{\max} = V(t)I(t)dt. \tag{33}$$

Therefore, according to Landauer's principle, one essential requirement has to be fulfilled to transfer the memristor device from state $\{w(t)\}$ to state $\{w(t + dt)\}$, namely

$$V(t)I(t)dt \geq n\,RTln2 \quad \text{on going from } \{w(t)\} \text{ to } \{w(t + dt)\}, \tag{34}$$

where $n$ denotes the mole number of the relevant species constituting the considered device material.

With regard to equation (34), the dynamical state equation (31b) would thus violate Landauer's principle as there is no restriction with respect to the minimum amount of energy which is needed to attain to an internal, physical state change. Following equation (31b), one would be able to change the state of a system – and that means in our case a real physical modification – at any time by merely feeding some electric current through the system, independent of the energy or work which can be actually supplied to the system in course of time. However, internal states of a system can only be altered if some minimum amount of generalized thermodynamic work is involved in, and that holds for both macroscopic and nanoscopic systems. Reasoning thoroughly about all of this, the dynamical state equation (31b) for our hypothetical memristor violates thus the fundamental requirements of the thermodynamics of information processing by itself. Maybe, approaches like the state equations (31a) and (31b) might be maintained if an extra side condition for the considered system is specified, namely the minimum electric power input to the device which is necessary to arrive continuously at internal, physical state changes, but we have reasonable doubts with regard to this. Physically, one might be confronted with capacitive or inductive effects, but it is beyond the scope of the present work to discuss such thinkable systems.

It is therefore time to consider alternative mechanisms to understand the variety of resistance switching phenomena which are observed on diverse material systems [25]. As we have shown, one has always to feed in a certain amount of energy to a material system in order to accomplish non-volatile resistance switching effects, *i.e.*, one has to produce some stable or metastable material modifications by surmounting Gibbs free energy barriers. As far as one is merely concerned with the redistribution of charged defects inside a single-phase material, one will never be able to yield stable resistance states; such systems would permanently "erase" all stored information by themselves as they always "strive" to reach at their absolute equilibrium states. Non-volatile resistance switching effects must thus result from either metastable physical modifications inside the considered material or from the impact of some imprinted electric polarization (ferroelectric resistance switching) on the overall current transport properties of the device.

With regard to material modifications, one could think of the precipitation of domain-like, metastable phase regions inside an otherwise homogeneous matrix. The redistribution of charged point defects due to an applied current or voltage stress can lead to strong, local aggregations of such defects into extended defects which finally cluster together to form a new phase when the stability limit of the main phase is exceeded [26].

Viewed under this aspect, TiO$_2$ would be a prototypical material [26] as it is characterized by a series of Magneli phases differing not much in Gibbs free energy so that the energy costs for achieving resistance switching would be manageable. However, to avoid that large resistance fluctuations crop up in such devices, the switching process has to be carried out in such a way that these metastable domains are large enough to be unsusceptible to thermal fluctuations [27]. We thus expect – when thinking of applications – that there will be restrictions with respect to the minimum dimensions of such devices.

## 6. Conclusions

The physical model presented by Strukov et al. [1] to describe the current-voltage relations of the HP-memristor device has been subjected to a thorough analysis. Based on a correct and valid physical description, we have shown that the proposed model suffers from a severe inconsistency. There are some fundamental facts of electrochemistry that have been overlooked by Strukov et al. [1] while putting forward their model. A "built-in" chemical inhomogeneity in single-phase TiO$_2$ results always in the appearance of a finite "polarization potential" as long as an oxygen vacancy concentration gradient exists inside the material. Following a valid analytical description, the HP-memristor device would act for a certain time like a "chemical capacitor" until the chemical inhomogeneity is balanced out. This, however, violates the "no energy discharge property" of a genuine memristor.

We could further show by analyzing the HP-memristor model in the framework of thermodynamics that such models must generally fail as the derived state equations violate by themselves Landauer's principle of the minimum energy costs in course of information processing. Maybe, such models might be maintained if an extra side condition is indicated for the related state equations to work, i.e., one would always have to specify the minimum electric power input which is necessary to continuously arrive at physical state changes of the device. With respect to this, however, we have reasonable doubts.

**Acknowledgements**

The authors are grateful to Felix Gunkel for many helpful discussions.

**Figures**

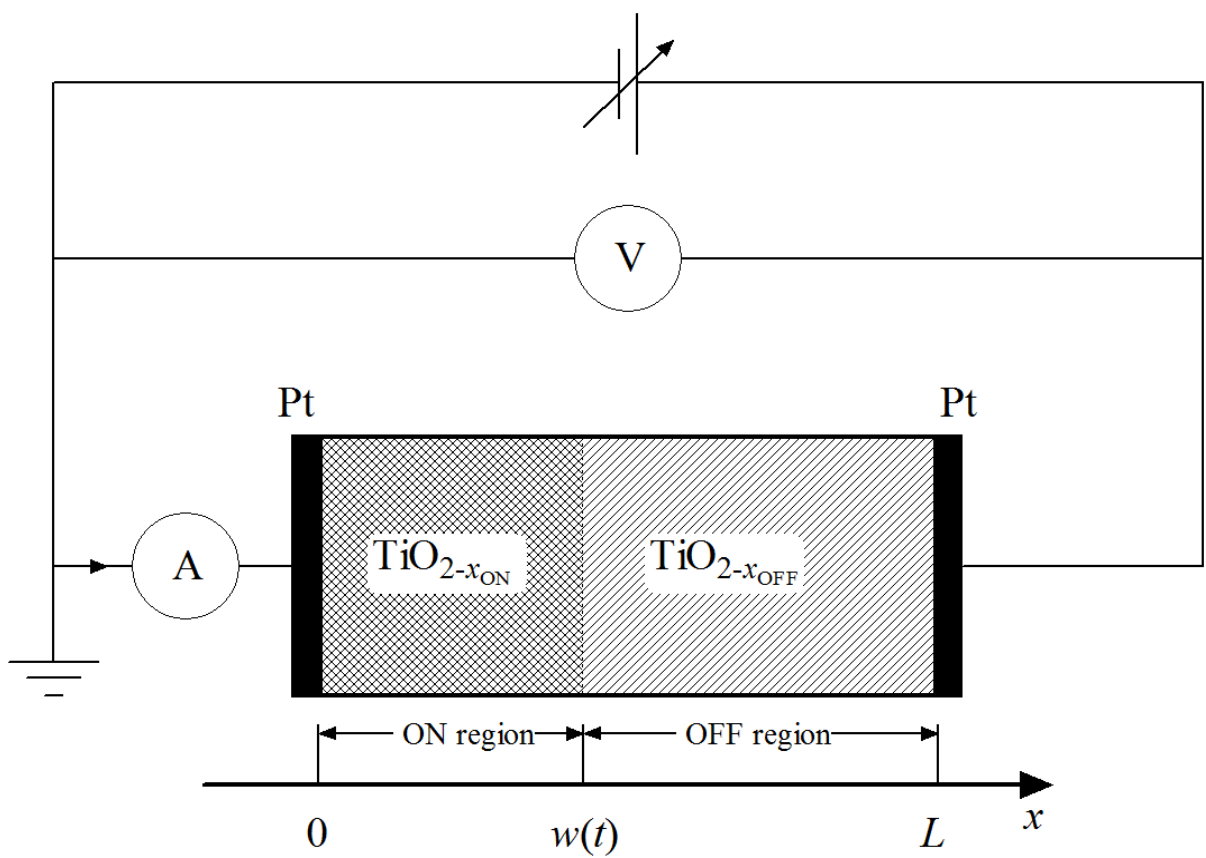

**Figure 1**. A schematic sketch of the HP-memristor device as a simplified equivalent circuit diagram